# An intertwined neural network model for EEG classification in brain-computer interfaces

Andrea Duggento[1], Mario de Lorenzo[2], Stefano Bargione[1], Allegra Conti[1], Vincenzo Catrambone[3,4], Gaetano Valenza[3,4], Nicola Toschi[1,5]

1. Department of Biomedicine and Prevention, University of Rome Tor Vergata, Rome (IT)
2. School of Biomedical Engineering, Science and Health Systems, Drexel University, Philadelphia (USA)
3. Bioengineering and Robotics Research Centre "E. Piaggio", University of Pisa, Pisa, Italy.
4. Department of Information Engineering, University of Pisa, Pisa, Italy.
5. Athinoula A. Martinos Center for Biomedical Imaging, Massachusetts General Hospital, Harvard Medical School, Charlestown, MA, USA

**Correspondence to:**
Dr. Andrea Duggento
Department of Biomedicine and Prevention
University of Rome Tor Vergata
Via Montpellier 1, 00133 Rome (IT)
duggento@med.uniroma2.it

# Abstract

The brain computer interface (BCI) is a nonstimulatory direct and occasionally bidirectional communication link between the brain and a computer or an external device. Classically, EEG-based BCI algorithms have relied on models such as support vector machines and linear discriminant analysis or multiclass common spatial patterns. During the last decade, however, more sophisticated machine learning architectures, such as convolutional neural networks (40, 41), recurrent neural networks, long short-term memory networks and gated recurrent unit networks, have been extensively used to enhance discriminability in multiclass BCI tasks. Additionally, preprocessing and denoising of EEG signals has always been key in the successful decoding of brain activity, and the determination of an optimal and standardized EEG preprocessing activity is an active area of research. In this paper, we present a deep neural network architecture specifically engineered to a) provide state-of-the-art performance in multiclass motor imagery classification and b) remain robust to preprocessing to enable real-time processing of raw data as it streams from EEG and BCI equipment. It is based on the intertwined use of time-distributed fully connected (tdFC) and space-distributed 1D temporal convolutional layers (sdConv) and explicitly addresses the possibility that interaction of spatial and temporal features of the EEG signal occurs at all levels of complexity. Numerical experiments demonstrate that our architecture provides superior performance compared baselines based on a combination of 3D convolutions and recurrent neural networks in a six-class motor imagery network, with a subjectwise accuracy that reaches 99%. Importantly, these results remain unchanged when minimal or extensive preprocessing is applied, possibly paving the way for a more transversal and real-time use of deep learning architectures in EEG classification.

# Introduction

The brain computer interface (BCI) (1-6) is a nonstimulatory direct and occasionally bidirectional communication link between the brain and a computer or an external device. BCIs were thought of as a viable technology for enhancing/replacing current neural rehabilitation techniques or providing assistive devices that were directly controlled by the brain (7-9). J. J. Vidal recorded the evoked electrical activity of the cerebral cortex from the intact brain in 1973, marking the beginning of a systematic endeavor to design an electroencephalogram (EEG, developed by Berger (10))-based BCI. The potential of BCI for improving human working capacity, both physically and cognitively, in subjects with, e.g., motor deficits (11, 12) was soon recognized, as it can aid in the artificial augmentation or re-excitation of synaptic plasticity in impacted brain circuits as an alternative to traditional rehabilitation therapy. While the underlying concept is to restore the connection between the brain and a compromised peripheral site by utilizing unaffected cognitive and affective functions (13), BCIs have also been used to control virtual reality (14), quadcopters (15), video games (16), and humanoid robots, as well as in other applications, including enterprise security (17), fingerprinting for lie detection, drowsiness detection for improving human working performance, estimating reaction time, and driving (18, 19). According to the European Commission, there are six main application themes: improve (e.g., upper limb rehabilitation after stroke), replace (e.g., BCI-controlled neuroprosthesis), enhance (e.g., enhanced user experience in computer games), supplement (e.g., augmented reality glasses), and research tool (e.g., decoding) (20).

Classically, BCI algorithms have relied on models such as support vector machines (SVMs) and linear discriminant analysis (LDA) or multiclass common spatial patterns CSP (21) and K-nearest neighbor (22). During the last decade, however, more sophisticated machine learning (ML) or deep learning (DL) architectures, such as convolutional neural networks (CNNs) (23, 24), recurrent neural networks (RNNs), long short-term memory (LSTM) networks and gated recurrent unit (GRU) networks (25), have been extensively used to enhance discriminability in multiclass BCI tasks. In the neuroprosthetic BCI field, a large effort has been devoted to simultaneous feedback-loop control of external devices through DL approaches. Different combinations of EEG-based experimental BCI paradigms (i.e., motor imagery for hands, (23, 26-28), fingers (29, 30), upper limbs (31), visual-evoked potentials (32, 33), cognitive-based BCIs (34, 35)) have demonstrated the potential of DL architectures as cutting-edge methods to recognize physical and mental states in both healthy and clinical populations. For instance, the use of a multimodal ML paradigm for EEG-engineered feature classification has supported healthcare practitioners in the biometric evaluation of dementia development (36), the forecasting of early schizophrenia risk insurgence in children (37), and screening for early Alzheimer's symptomatology (38). Likewise, AI-driven neurofeedback with BCI has been employed in motor imagery (MI) rehabilitation training and specifically in the functional modulation of the brain's sensorimotor rhythms (39).

Preprocessing and denoising of EEG signals has always been key in the successful decoding of brain activity. In this context, deep architectures such as generative adversarial networks (GANs) have been employed as part of an automatic pipeline for single-subject EEG signal denoising (40). In comparison to algorithms typically applied for medical diagnostic purposes (41), end-to-end architectures allow the incorporation of EEG signal preprocessing stages into the recording phase, hence paving the way for more flexible real-time applications that include a better representation of subjectwise variability and more adaptability in a personalized medicine context. Importantly, a major hurdle in propelling BCI architectures beyond the research environment and into clinical practice is often hampered by the fact that most algorithms currently in use are based on preemptive feature extraction and therefore cannot be employed in real time.

In this paper, we present a deep neural network architecture specifically engineered to a) provide state-of-the-art performance in multiclass MI classification and b) remain robust to preprocessing to enable real-time processing of raw data as it streams from EEG and BCI equipment.

## Materials and Methods

### Dataset description

We made use of the halt motor imagery paradigm data made available in (42). In brief, 12 subjects (originally labeled in (42) from 'A' to 'M', where subject 'D' was missing) underwent 3 sessions of EEG recording each while performing a motor imagery task when a visual cue was provided through an eGUI. Each session consisted of a series of BCI interaction trials. During each trial, a 1-second-long visual cue prompted through the eGUI instructed the participants to implement a given mental image, followed by a random duration (1.5 to 2.5 s). On average, each trial was repeated approximately 900 times per session, where each session was split into 3 segments of 15 min each to allow the participant to recover from mental fatigue. During rest, participants were asked to look at a central fixation point. At the beginning of each trial, an action signal indicating the left hand, right hand, left leg, right leg, tongue or a circle (indicating passive response) was presented (see Fig. 1).

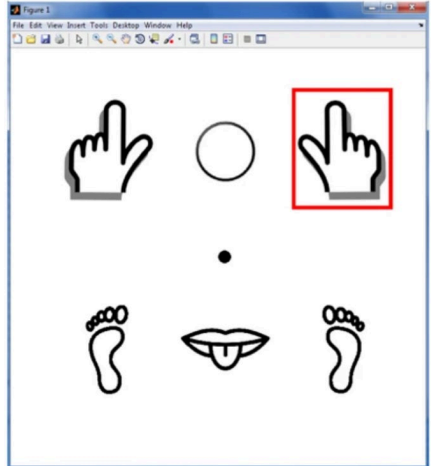

**Figure 1:** *Depiction of the eGUI during the halt interaction paradigms. Around a passive image, the eGUI displays five icons that indicate motor imagery (left hand, right hand, left leg, right leg, or tongue) around a passive image (fixation circle, center). The cue for interaction for any given task was given by showing a red rectangle around each icon task. Adapted from (59).*

Subjects were instructed to mentally depict a brief movement of either the left or right-hand or left or right-foot, to imagine a movement of the tongue that would imply pronouncing a specific letter or sound (such as "el."), or to remain mentally 'passive', i.e., not to imagine any movement. This resulted in six classes among which to discriminate.

### Data acquisition and preprocessing

In (59), EEG data were acquired with 19 bridge electrodes in the 10/20 international configuration at a sampling frequency of 200 Hz. An in-hardware bandpass filter of 0.53-70 Hz was originally present in all EEG data. Then, we generated two versions of the dataset: in one, we applied a bandpass filter at 8-30 Hz since it is widely believed that this range of frequencies contains most MI information, with no other preprocessing (43). In the other, EEG data were extensively preprocessed using part of the Harvard Automated Processing Pipeline for Electroencephalography (HAPPE) (44). In brief, this entailed 0.5-70 Hz bandpass filtering, line noise rejection, session segmentation, common average re-referencing, crude bad channel detection using spectrum criteria and three standard deviations as channel outlier threshold, wavelet-enhanced ICA (where ICA was first used for clustering the data, then wavelet decomposition was employed for thresholding the ICs), and interpolation of the channels that were flagged as bad earlier. The two versions of the datasets are intended to explore the impact on our classification pipeline of extended, established preprocessing steps versus minimally preprocessed data, which is more apt for real-time BCI applications.

**Architectures**

To illustrate the rationale that underpins the novel architecture presented in this paper, we first discuss the baseline architectures that serve as baselines as well as their assumptions.

*Cascade and parallel architectures*

In (45), Zhang et al. proposed two architectures, termed 'cascade' and 'parallel', composed of convolutional and recurrent neural layers organized either one after another (cascade architecture) or simultaneously processing the same data in space and in time (parallel architecture). The aim was to exploit both spatial correlations between physically neighboring EEG sensors and temporal dependencies with an array of LSTM layers. The cascade network, depicted in Fig. 2, extracts (at every time point) complex spatial patterns with the convolutional layers and then pipes the obtained latent representation into layers of LSTM to embed subtle time dependencies of the time-varying latent representation.

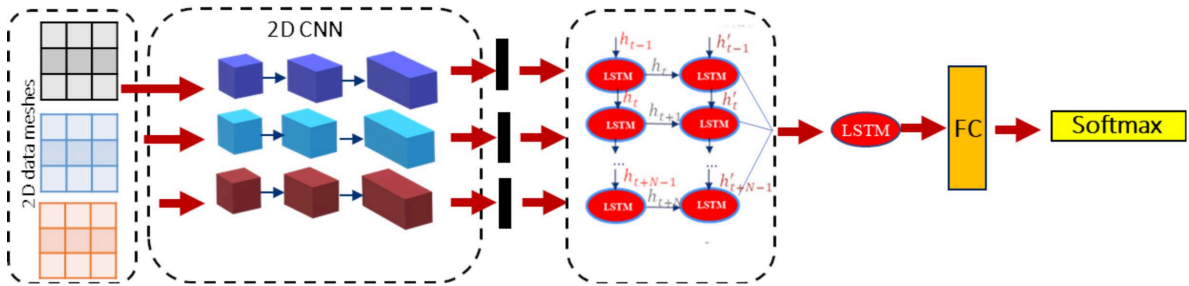

**Figure 2:** *Representation of the cascade architecture depicting convolutional and recurrent neural network layers whose output is piped.*

Conversely, the parallel network, depicted in Fig. 2, extracts complex spatial patterns *and* time dependencies from the EEG data and merges them with layers of fully connected neurons.

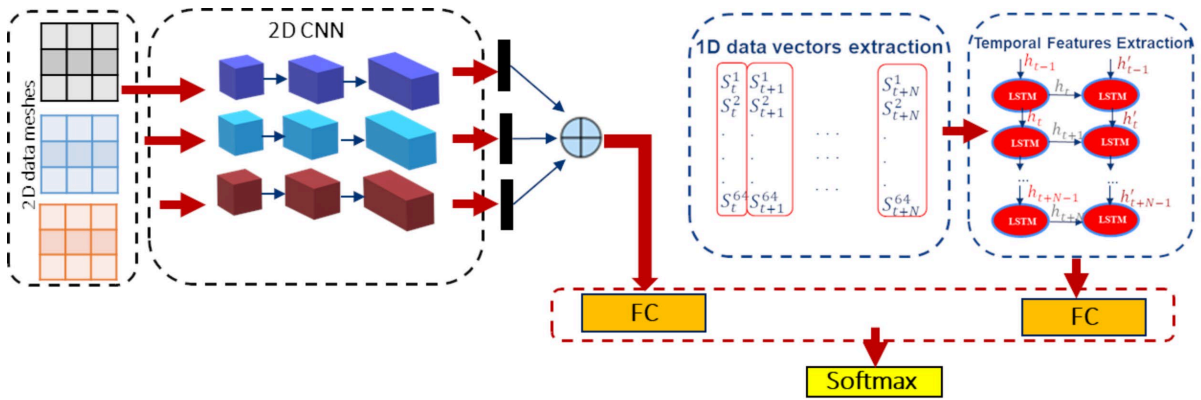

**Figure 3:** *Representation of the parallel architecture depicting the recurrent convolutional neural layer and convolutional layers in parallel and concatenated before fully connected layers*

The cascade and parallel networks reflect two different ideas about how to tackle the EEG classification problem and develop two different rationales. The cascade network assumes that, irrespective of time-structure, sophisticated spatial patterns emerge, which are transformed into a latent space-representation, which, in turn, must be 'followed' in time with recurrent layers. Instead, the parallel network assumes that while a spatial complex pattern emerges, every signal from each EEG electrode contains an intricate spatial pattern that is best decoded in a separate latent temporal-representation, which, in turn, will be merged with the latent space-representation. Hence, both cascade and parallel networks assume a certain degree of separation between the nonlinear spatial and temporal information 'manipulations'.

*Intertwined architecture*

In this paper, we introduce an architecture to explicitly address the possibility that interaction of space features and temporal features occurs at all levels of complexity: simple spatial features evolve in time, whose time evolution can be represented by other, more complex, space features, which evolve in time. Our proposed architecture, which we term the *intertwined* architecture, is composed of alternated time-distributed fully connected layers (tdFC) and space-distributed 1D time-convolutional layers (sdC), followed by several LSTM layers and, successively, FC layers, before the last layer is employed for classification (six neurons, corresponding to the six classes present in the dataset).

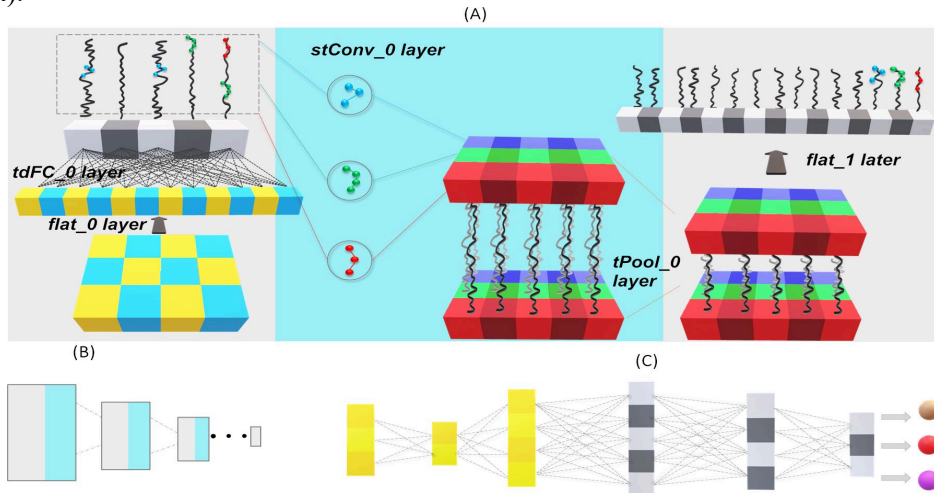

**Figure 4:** *Representation of the intertwined architecture: (A) the basic module consists of a flattening layer, a tdFC layer, a sdC layer, and a pooling layer. (B) Basic modules can be concatenated, and (C) the output of the last module is the input to the LSTM and FC layers.*

Let us assume that the shape of the input data is $L \times K$, where $L$ is the number of electrodes and $K$ is the number of time points in each trial (in our case, $19 \times 200$). Fig. 4(A) shows the details of the first intertwined module: the $L \times K$ time-signals coming from all electrodes are flattened along the space dimension; then, the first tdFC layer composed of $tdFCnum_0$ neurons is fully connected and time-distributed (meaning that the weights and biases are the same for all time steps). After a time-distributed neuron's activation function (tdNact), we obtain a $tdFCnum_0 \times K$ signal that is connected to a sdC layer (weights and biases are the same for all $tdFCnum_0$ components) composed of $sdCnum_0$ kernels, with size *sdCker* (stride=1) each and with a sdCact. After the sdC layer and the relative sdCact, we obtained a $tdFCnum_0 \times sdCnum_0 \times (K - sdCker - 1)$ output. The last dimension, which is the time occurrence of the actvation of the time-convolutional kernels, is then 'compressed' with a pooling layer called $tPool_0$, while the first two dimensions, which contain the interaction between the space features and the time kernels, are flattened. After the first intertwined module, the output, like the input to the input, contains a 'space-like' dimension and a time-dimension and is ready to be processed by another intertwined module. Fig. 4(B) shows the concatenated intertwined modules, and Fig. 4(C) shows that the end of the last intertwined module, the output, is piped to several LSTM layers and then to *FC* layers.

*Hyperparameter search*

For every subject and for every architecture, we conduct an extensive search in the hyperparameter space that defines the architecture while minimizing the cross entropy for the 6-class classification task.

**Cascade architecture**

We optimized the number {16, 24, 48, 92}, the size {2, 3}, and the stride {1, 2} of convolutional kernels as well as their number of layers {1, 2, 3}; the number of LSTM neurons {10, 50, 100} and the number of LSTM layers {0, 1, 2, 3}; the number of layers {0, 1, 2} of FC neurons and the number of FC neurons per layer {10, 50, 100} after both CNN-branch and LSTM-branch.

**Parallel architecture**

We optimized the number {16, 24, 48, 92}, the size {2, 3}, and the stride {1, 2} of convolutional kernels as well as their number of layers {1, 2, 3}; the number of LSTM neurons {10, 50, 100} and the number of LSTM layers {0, 1, 2, 3}; the number of layers {0, 1, 2, 3} of FC neurons and the number of FC neurons per layer {10, 50, 100}.

**Intertwined architecture**

Given the relatively high number of hyperparameters, the search was conducted with a few discrete values per parameter. We optimized the following hyperparameters: **modNum**: Number of intertwined modules (2,3, 4); **$tdFCnum_i$**: Number of tdFC neurons; independently varied for each $0 \leq i < modNum$, defines the number of tdFC neurons (16, 24, 50, 80). **tdNact**: activation type for tdFC neurons. (relu, softmax, elu, selu). **$sdCnum_i$**: independently varied for each $0 \leq i < modNum$, defines the number sdC kernels (16, 24, 50, and 80). **$sdCker_i$**: independently varied for each $0 \leq i < modNum$, defines the size of each sdC kernel (2, 3, 4, 5). **sdCact**: activation type for sdC neurons (relu, softmax, elu, selu). **poolSize**: time-pooling size after each intertwined modul (2, 3, 4). **pooltype**: time-pooling type after each intertwined module ('MaxPooling', 'AveragePooling'). { **$LSnum_i$**}: vector of integers, independently varied for each $i$, defines the number of LSTM neurons in each layer. If $LSnum_0 = 0$, then no LSTM layers are employed, and the time dimension is pooled with a 'GlobalAverage' pooling layer (0, 30, 50, 100, 200). **LSdrop**: dropout coefficient that determines the dropout ratio in the LSTM layer connections (fixed at 0.1). { **$FCnum_i$**}: vector of integers, independently varied for each $i$, defines the number of neurons in each FC layer. If $FCnum_0 = 0$, then no FC layers are employed with the exception of the last 6-neurons-for-6classes classification layer {0, 30, 50, 100}. **FCact**: activation type for FC neurons {relu, softmax, elu, selu}. **FCdrop**: dropout coefficient that determines the dropout ratio in FC layer connections (fixed at 0.1). **minimizer**: type of minimizer employed for training the architecture {*sgd, rmsprop*}.

# Results

The accuracies of the best performing model on each subject are listed in Table 1 and summarized in

Fig 5. To statistically compare results across model families, we employed a related samples two-way Friedman's analysis of variance (a nonparametric test for repeated measures) across model families, which returned an overall p-value of 0.002. Subsequent pairwise comparison (adjusted across comparisons using Bonferroni's correction) showed no significant difference in accuracy across subjects between the cascade and parallel model families (p=0.124), while the intertwined model returned significant differences in accuracy both when compared to the parallel (p=0.024) and cascade (p=0.001) model families, with the intertwined model family showing the highest accuracy overall.

| Subject | Intertwined | Parallel | Cascade |
|---------|-------------|----------|---------|
| A | 0.84 | 0.75 | 0.33 |
| B | 0.48 | 0.36 | 0.26 |
| C | 0.93 | 0.82 | 0.47 |
| E | 0.77 | 0.58 | 0.26 |
| F | 0.59 | 0.49 | 0.30 |
| G | 0.89 | 0.81 | 0.49 |
| H | 0.38 | 0.28 | 0.20 |
| I | 0.59 | 0.35 | 0.38 |
| J | 0.99 | 0.96 | 0.82 |
| K | 0.85 | 0.67 | 0.65 |
| L | 0.96 | 0.85 | 0.66 |
| M | 0.82 | 0.63 | 0.52 |

**Table 1:** *Subjectwise validation accuracies obtained at best validation loss on 12 subjects over a 6-label task. We optimized the hyperparameters for each model family and for each subject. For each subject, for each model family, the validation accuracy was derived from the best (lower loss function)-performing model.*

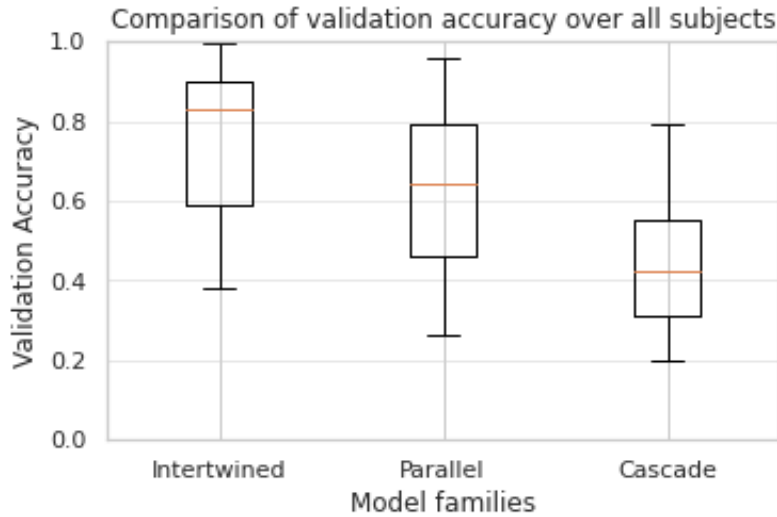

**Figure 5**: *Comparison of validation accuracy obtained at best validation loss on 12 subjects over a 6-label task. For each model family and for each subject, a massive sweep with respect to model hyperparameters was carried out. For each subject, for each model family, the validation accuracy was derived from the best (lower loss function)-performing model. Median accuracies between model families were significantly different (p=0.002, Friedman's analysis of variance).*

**Sensitivity to hyperparameters**

In addition to structural hyperparameters that define the architecture shape (numbers of layers, numbers of neurons, etc.), we observed a marked dependency in both the performance (minimum validation loss) and training time (epoch at which such a minimum is reached) with respect to two key variables. The first variable is the activation function after every fully connected layer and after every time-convolutional layer. A total of four activations were tested: *relu*, *softmax*, *elu* and *selu*. Exemplary results are depicted in Fig. 6.

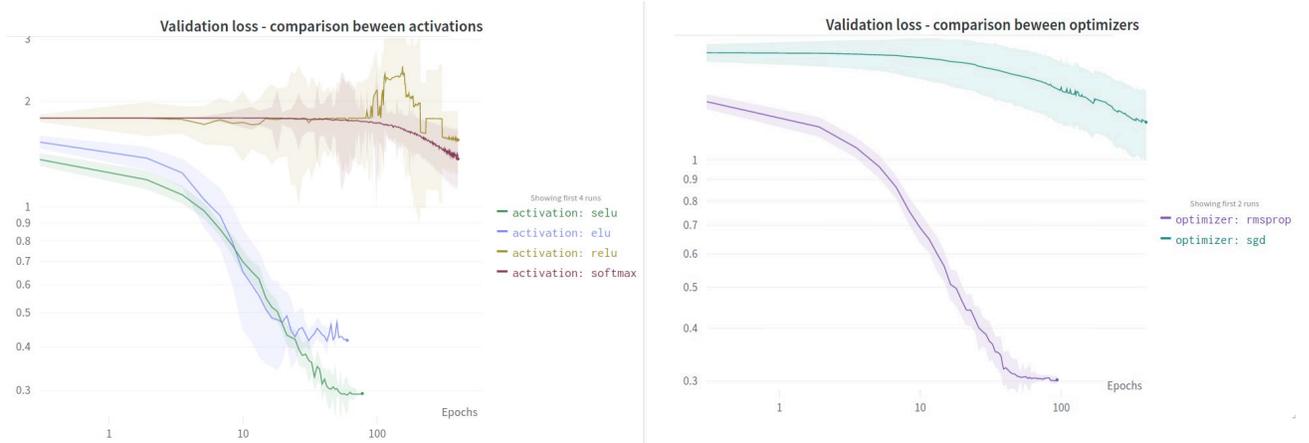

*Figure 6*: Left: Exemplary comparison of validation loss with respect to epoch number by using four different activation functions: Right; Exemplary comparison of validation loss with respect to epoch number by using two different optimizers. The results shown refer to subject 'G' (randomly chosen). Solid lines: median validation loss of the best 50 runs; shaded areas: standard deviation of the best 50 runs.

The second variable is the **minimizer**. Two separate minimizers were employed: sgd and rmsprop. Models with *elu* and *selu* activations largely outperform models with relu and softmax activations, both in training speed and in overall performance, with *selu* performing marginally better than *elu* with respect to overall performance. Figure 6 also clearly shows that the *rmsprop* optimizer vastly outperformed the *sgd* optimizer in our model and implementation.

**Impact of preprocessing**

The presence or (almost complete) absence of preprocessing did not significantly impact the accuracy of our intertwined architecture (see Table 2 and Figure 7).

| Subject | Minimal | Extensive |
|---|---|---|
| A | 0.84 | 0.85 |
| B | 0.48 | 0.46 |
| C | 0.93 | 0.92 |
| E | 0.76 | 0.76 |
| F | 0.59 | 0.63 |
| G | 0.89 | 0.92 |
| H | 0.38 | 0.37 |
| I | 0.59 | 0.57 |
| J | 0.99 | 0.99 |
| K | 0.85 | 0.83 |
| L | 0.96 | 0.96 |
| M | 0.82 | 0.8 |

*Table 2:* *Comparison of validation accuracy obtained at best validation loss on 12 subjects over a 6-label task. For each subject, the validation accuracy was derived from the intertwined architecture trained on the minimally vs extensively preprocessed EEG*

**Figure 7: Left:** Box-Whisker plots (across subjects) of validation accuracy obtained at best validation loss from the intertwined architecture trained on the minimally vs extensively preprocessed EEG data *(p=0.96, Friedman's analysis of variance).* **Right:** Subjectwise accuracy data for the two preprocessing strategies.

# Discussion

In this paper, we have described a novel deep learning architecture that provides state-of-the-art performance in discriminating multiclass EEG on a recent and well-curated motor imagery dataset. In contrast to previously presented deep learning paradigms that handle the spatial and temporal complexity as well as interrelation intrinsic to the EEG signal in separate blocks, we exploit the idea of creating spatiotemporal representations that interact at all architectural and complexity levels. This allows for greater representational ability as well as for a more natural description of the nature of actual EEG data. When benchmarking our method against a recent, extremely successful architecture that combines LSTM and CNN models in different arrangements (45), we obtain a statistically significant increase in performance, with accuracies as high as 99% in single subject classification, also outperforming all approaches recently summarized in an extensive comparative study performed on the same dataset (46). At the same time, we observe a very large intersubject variability in the maximum attainable performance (at best loss value), which is mirrored in the other methods we employed in this paper. While this means that a universally optimal BCI model is not yet within reach, it is also well known that interindividual variability plays an extremely large role in BCI performance (47), and it is likely that the integration of other physiological signals may be necessary to achieve stable, high-end accuracy in multiclass problems across different subjects.

EEG preprocessing, especially when working at the single-trial level, as in this paper, is commonly considered crucial to separate physiologically meaningful information from electronic and/or physiological noise in EEG. In this context, the effect of preprocessing techniques on the effectiveness of deep learning models has recently been investigated (48) in mental workload tasks. The authors compared numerous automated preprocessing pipelines similar to the one employed in this study and concluded that preprocessing steps such as the ADJUST algorithm (49) in conjunction with other steps such as common rereferencing had a differential impact on classification performance when using stacked LSTM, bidirectional LSTM (BLSTM), and BLSTM-LSTM models. In contrast, our model showed virtually no difference in performance when employed with minimally or extensively preprocessed data. In this context, there is an ongoing discussion on how EEG data and related computation can fit into the modern personalized medicine paradigm (50). Standardizing medical practices and methodologies can enhance the efficacy and security of medical interventions, and to contribute to what appears to be the future path of modern medicine, several authors advocate a transition to the field of quantitative EEG (qEEG), in which individuals' deartifacted resting state EEGs are compared to a normative database to assess clinically significant deviations. While this is certainly a crucial aspect of employing biosignals as modern biomarkers, the apparent resilience of our model to artifacts would, in principle and when validated on larger and more variable datasets, provide an alternative strategy to ensure intrasubject robustness. We speculate that this feature is due to the absence of convolutional blocks, making our model substantially agnostic to the number and positioning of the electrodes. This contrasts with other modern EEG classification architectures that assume a certain degree of spatial regularity, which, however, does not scale linearly with the number of electrodes, potentially hampering applicability in modern high-density EEG applications.

When exploring hyperparameter values for our intertwined architecture, we discovered a significant advantage in using the *selu* and to a lesser estent, *elu* activation functions, as well as of the *rmsprop* optimizer, which is often employed in recurrent neural networks such as the ones employed at various stages of our architecture. These empirical results point toward the importance of self-normalization for processing EEG data and could provide important clues that can generalize to any DL architecture aiming at EEG data modeling.

## Conclusions

We present a novel DL architecture for EEG classification that is agnostic to electrode positioning and outperforms state-of-the-art deep architectures based on combinations of convolutional and recurrent neural network blocks. The classification results on a recently published, high-quality 6-class motor imagery dataset are insensitive to preprocessing, possibly bypassing a major source of arbitrariness and paving the way for wider, bedside implementations of EEG-based BCI applications.

## List of abbreviations

MI: motor imagery
BCI: brain computer interface
EEG: electroencephalography
ReLU: rectified linear unit
softmax: softmax
ELU: exponential linear unit
SELU: scaled exponential linear unit
SGD: stochastic gradient descending
RMSProp: root mean squared propagation
HaLT: hands-legs tongue
eGUI: electronic graphical user interface
tdFC: time-distributed fully connected layers
sdConv: space-distributed 1D time-convolutional layers
LSTM: long-short term memory
FC: fully connected
tdNact: time-distributed neuron activation function
sdCact: space-distributed time-convolutional activation function

## Acknowledgments
This project was supported by the EXPERIENCE project, funded by the European Union's Horizon 2020 research and innovation programme under grant agreement No. 101017727. Stefano bargione is a PhD student enrolled in the National PhD in Artificial Intelligence, XXXVII cycle, course on Health and life sciences, organized by Università Campus Bio-Medico di Roma.

## Code availability
The code employed in this paper to run the intertwined architecture is available at https://github.com/andreaduggento/EEG_intertwined_architecture